\def\morethan#1{{ \it et al.}}

\def\alwaysmath#1{\ifmmode{#1}\else{$#1$}\fi}

\documentclass[12pt,preprint]{aastex} 
 
\slugcomment{to appear in Astrophysical Journal} 
 
\shorttitle{Solving the $\omega$ Cen  enigma} 
\shortauthors{Ferraro, Bellazzini \& Pancino} 
 
\begin{document} 
 
 

\title{Discovery of an  accreted stellar
system within the globular cluster  $\omega$ Cen\thanks{Based  on WFI 
observations collected at La Silla, Chile, within the observing 
programmes 62.L-0354 and 64.L-0439.}} 
 

\author{F. R. Ferraro, M. Bellazzini} 
\affil{Osservatorio Astronomico di Bologna, Via Ranzani 1, I-40127, Bologna,  
Italy.} 
\email{ferraro@apache.bo.astro.it, bellazzini@bo.astro.it} 
  
\and 
\author{E. Pancino\altaffilmark{2}} 
\affil{European Southern Observatory, K. Schwarzschild Str. 2,  
Garching, D-85748, Germany.} 
\email{epancino@eso.org}

 
\altaffiltext{1}{on leave from Dipartimento di Astronomia, 
Universit\`a di Bologna, Via Ranzani 1, I-40127, Bologna, Italy.}

\begin{abstract} 
Recent wide-field photometric surveys (Lee et al.; Pancino et al.)
have shown the existence of a previously unknown metal-rich
($[Fe/H]\simeq -0.6$) stellar population in the galactic globular
cluster $\omega$ Centauri.  The discovery of this new component, which
comprises only a small percentage ($\sim 5\%$) of the entire cluster
population, has added a new piece to the already puzzling picture of
the star formation and chemical evolution of this stellar system.  In
this Letter we show that stars belonging to the newly discovered
metal-rich population have a coherent bulk motion with respect to the
other cluster stars, thus demonstrating that they formed in an
independent self-gravitating stellar system.  This is the first
clear-cut evidence that extreme metal-rich stars were part of a small
stellar system (a satellite of $\omega$ Centauri?) that has been
accreted by the main body of the cluster. In this case, we are
witnessing an {\em in vivo} example of hierarchical merging on the
sub-galactic scale.
\end{abstract}    
 
\keywords{globular clusters: individual ($\omega$~Cen)---stars: kinematics 
---stars: Population II} 
 
 
\section{Introduction} 
\label{intro}

The giant globular cluster $\omega$~Centauri is the only Galactic
globular that shows multiple stellar populations, with a very
different heavy metal content. Large photometric
\citep{lee,hk00,hughes,p00} and spectroscopic \citep{nor96,smith}
surveys have demonstrated the presence of a main population at
${\rm[Fe/H]}=-1.6$ with a metal-rich tail extending up to
${\rm[Fe/H]}=-0.6$. The detailed abundance patterns of tens of red
giants spanning the entire metallicity range have been derived
\citep[see][]{norris,smith,p02}. The cluster also shows dynamical
anomalies with respect to ordinary globulars, since it is partially
substained by rotation \citep{may97,merritt}. Moreover, while the
metal-poor giants (${\rm[Ca/H]}<-1.2$) do rotate, the metal-rich
ones do not show any sign of rotation \citep[][see below for further
discussion]{nor97}.

Since its vicinity allows a very detailed study of its stellar
content, this giant cluster represents a cornerstone in our
understanding of the formation, the chemical enrichment and the
dynamical evolution of  stellar systems. However, in
spite of the huge observational effort carried out so far, the global
scenario for the formation and evolution of $\omega$~Cen still remains
a mystery \cite[see for instance][and discussions
therein]{nor96,smith}. To explain its peculiar properties and its
heterogeneous mix of populations, different hypotheses have been
suggested: {\it (i)} $\omega$~Cen is the relic of a larger galaxy
partially disrupted by the tidal field of the Milky Way
\citep{freeman,lee,maj}; {\it (ii)} it is the result of the merging of
two globulars \citep{searle,icke,makino}.  However, while some hints
of a complex dynamical history have been found \citep{nor97,jur}, none
of the evidence presented so far is conclusive, and most important,
the simple model of the merging of two ordinary globulars cannot fit
the observed broad metallicity distribution \citep{nor96,smith}.

\subsection{The sub-populations of $\omega$ Centauri} 
\label{sub-pops}

Recent wide-field photometric studies have added a new piece to this
already complex jigsaw puzzle: an additional, well-defined Red Giant
Branch (RGB) sequence has been discovered to the red side of the main
RGB in the Color Magnitude Diagram \cite[CMD;][]{lee,p00}.  The
discovery of this anomalous branch (hereafter RGB-a) unveiled the
existence of a further, previously unknown, metal rich sub-population
in $\omega$~Centauri.  In particular \citet{p00}, taking advantage of
the high accuracy and sensitivity of their $(B,B-I)-$CMD, have
identified three main sub-populations on the basis of their
photometric properties: the main metal-poor component (MP;
${\rm[Fe/H]}<-1.4$); the metal-intermediate population (MInt;
$-1.4\le{\rm[Fe/H]}<-0.8$) and the anomalous population (RGB-a) which
represents the extreme metal-rich end of the observed metallicity
distribution.

\begin{figure}    
\plotone{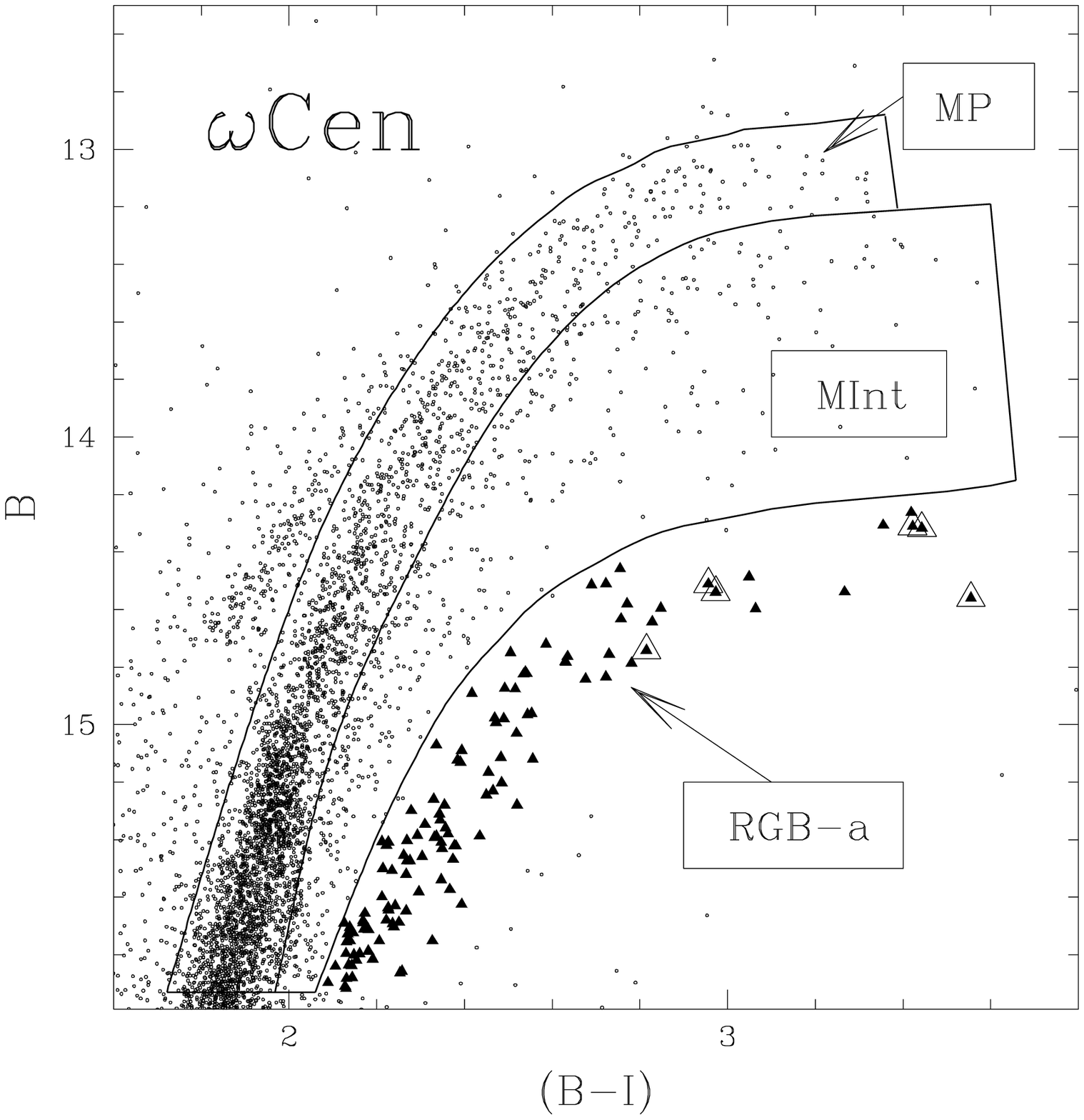}    
\caption{
Approximate locations of the three RGB sub-populations (defined in the
text) in the CMD, RGB-a stars are plotted as small solid triangles.
The six RGB-a stars which have been identified and studied before the
discovery of the anomalous branch are highlighted by large empty
triangles.
\label{wfi-cmd}}    
\end{figure}

The approximate CMD locations of the three sub-populations defined
above are shown in Figure~\ref{wfi-cmd}.  Stars belonging to the RGB-a
are plotted as {\it small solid triangles}.  Note that only a few
stars belonging to this population were known before the discovery of
the anomalous branch: in particular, only 6 RGB-a stars 
(marked as {\it
large open triangles} in Figure~\ref{wfi-cmd})
have previous
metal abundance determinations \citep{evans,nor96} and only 8 have
radial velocity determinations \citep{may97}.

Though the RGB-a population comprises only a small fraction
\citep[$\sim5\%$;][]{p00} of the cluster's stellar content,
its origin and evolution is the new challenging question in the
picture of $\omega$~Cen formation and evolution.
 
As part of a long-term project \citep{post} devoted to reconstruct the
global evolutionary history of $\omega$~Centauri, we are building a
comprehensive catalog combining all the relevant information available
in the literature for stars in this cluster. Here we present the first
results obtained by cross-correlating our wide field, multi-band
photometric catalog \citep{p00}, comprising more than 220,000 stars,
with the recently published proper motion study for $\sim 10,000$
stars brighter than $V\sim16.0$ \citep{vanl00}. The synthesis of these
data-sets allows us to study for the first time the kinematical
properties of the RGB-a stellar population.

\section{Results} 
 
\begin{figure}    
\plotone{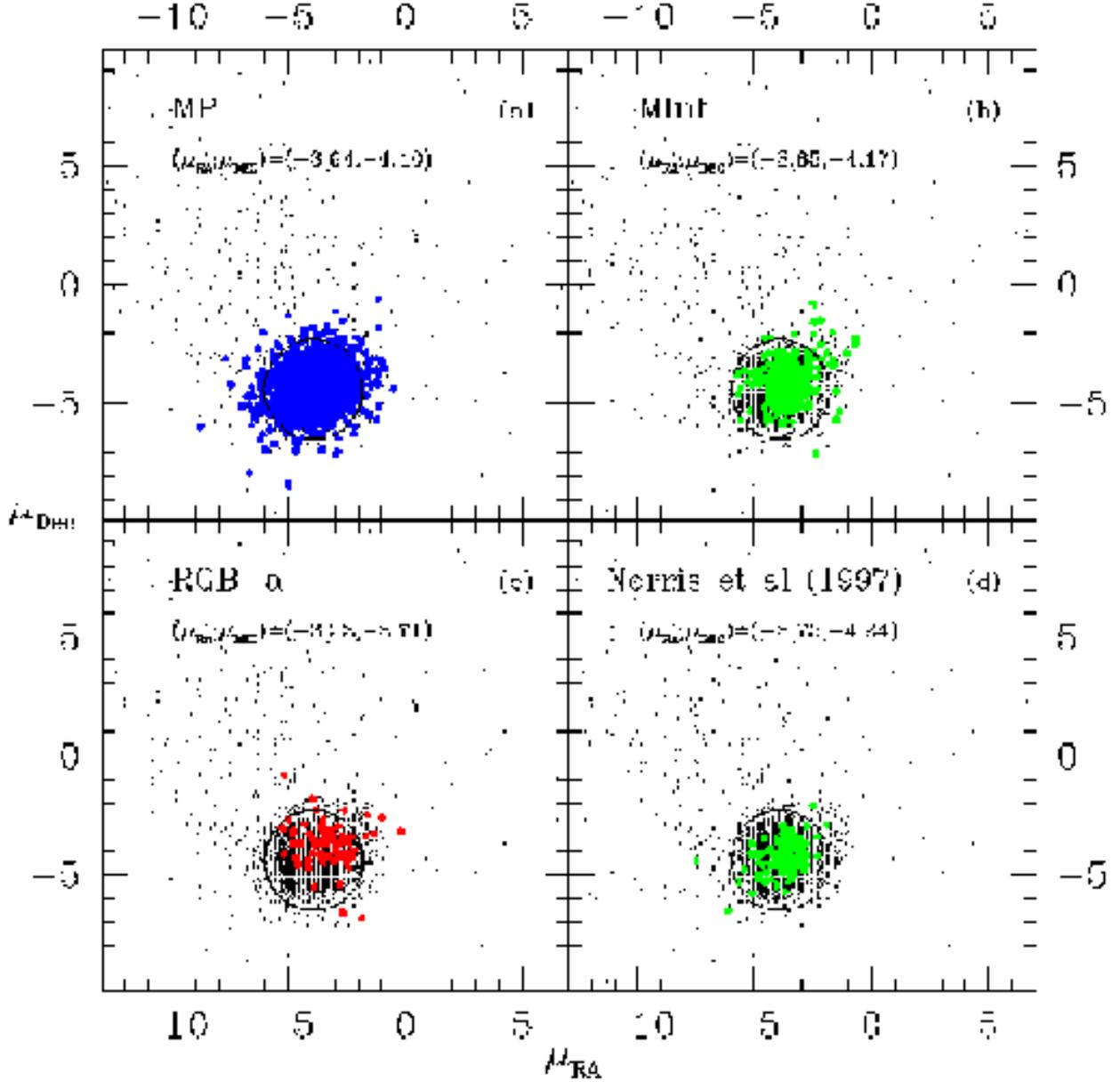}    
\caption{Proper motion plane $(\mu_{RA},\mu_{Dec})$ for stars    
in $\omega$ Cen. All stars in the Van Leween's catalog with accurate
proper motion measures have been plotted ({\it small dots}). Only
stars with accurate measures ($\sigma_{\mu}<0.5$ mas/yr) and {\it
quality flag}$< 4$ have been considered.  The solid circle deliniates
the region where stars with a membership probability larger than $99
\%$ lie. In the first three panels stars belonging to the three RGB
sub-populations defined in the text (see Figure 1) are marked with
filled dots: the MP, the MInt and RGB-a populations in ({\it (a), (b)}
and {\it (c)}, respectively.  The dashed circle in {\it panel (c)}
highlighted the center of symmetry for the RGB-a population.  The
proper motion distribution for the {\it `` metal-rich''} sub-sample
($[Ca/H]<1.2$) discussed by Norris et al (1997) is shown in {\it panel
(d)}, for comparison.  As can be seen, the Norris et al sample is
fully consistent with the MInt population and does not show the
coherent residual proper motion signature detected in the RGB-a stars.
The proper motion of the centroid for each population is reported in
each panel. In addition, it is interesting to note that the three
sub-populations show a similar dispersion $(\sigma_\mu\sim0.6$
mas/yr).
\label{motion}}    
\end{figure}

Figure~\ref{motion} shows the proper motion distributions for the
three different sub-populations defined above (see
Figure~\ref{wfi-cmd}): MP, MInt, and RGB-a (panels {\it (a)}, {\it
(b)} and {\it (c)}, respectively).  To exclude any effect due to
uncertain measures we consider here only the stars with safe and
accurate proper motions in the Van Leuween's catalog (i.e. quality
flag $<4$ and $\epsilon_\mu <0.5$ mas/yr).  For sake of comparison,
all the stars satisfying these selection criteria have also been
plotted in each panel as {\it small dots}.

The most striking result is that the proper motion centroid of the
RGB-a group is quite different from that of the MP population. This
means that the RGB-a stars have a coherent residual proper motion with
respect to the dominant cluster population. The systematic residual
turns out to be $\delta\mu_{RA}=+0.4$ and $\delta\mu_{DEC}=+0.7$
mas/yr, which corresponds to a total proper motion modulus of
$|\delta\mu_{TOT}|=0.8$ mas/yr, with respect to the centroid of the MP
population.  In order to exclude that the effect found above is due to
any kind of systematic error (for example to spurious effects on the
proper motion measures correlated to the color or the magnitude of the
stars) we also considered the proper motion distribution of the
Horizontal Branch (HB) stars, which have quite different colours and
luminosities with respect to the giants: no systematic difference
could be detected between the proper motion distributions of the HB
and the RGB-MP groups. A two-dimensional generalization of the
Kolmogorov-Smirnov test \citep[KS;][]{fasano}, applied to the proper
motion distributions of the HB and MP (see Table~\ref{ks}), suggest
that the two samples are fully consistent, i.e., they are extracted
from the same parent distribution.  This test indicates that the
systematic proper motion residual found for the RGB-a stars cannot be
ascribed to any spurious colour/luminosity effect.

\begin{figure}    
\plotone{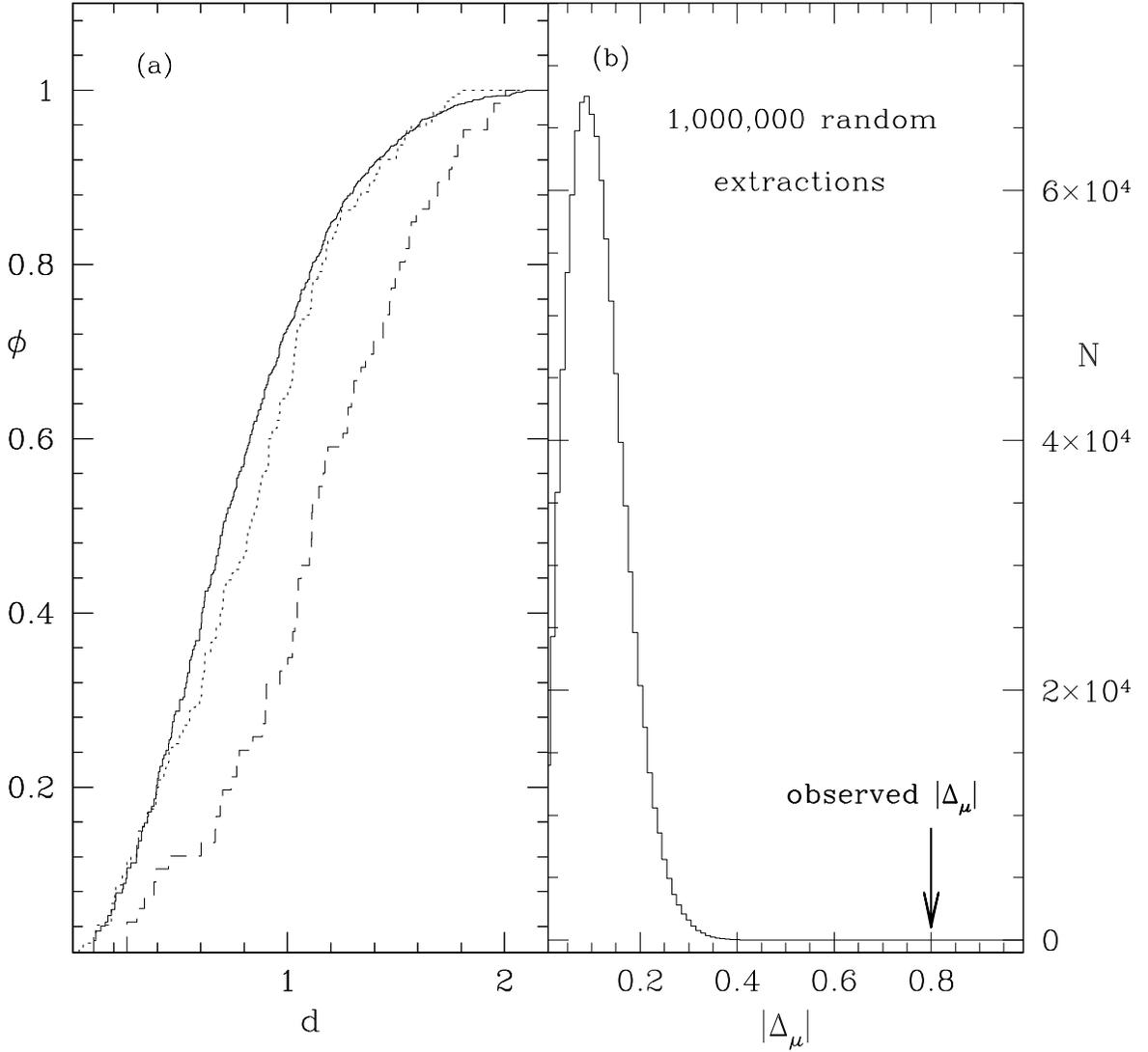}    
\caption{{\it Panel (a)}: Cumulative distribution ($\phi$)    
in the proper motion plane (see Figure 1), for the three 
sub-population (MP: {\it heavy solid line}; MInt {\it dotted 
line}; RGB-a {\it dashed line}). The X-axis variable ({\it d}) 
is the geometric 
distance of each star to the MP proper motion centroid
($(\mu_{RA}=-3.94,\mu_{DEC}=-4.40)$ in units of mas/yr).
Only stars with  accurate  
  proper motion measurements (see 
  the caption of Figure 1) and lying within the circle shown
 in {\it panel (a)}
of Figure 1 have been considered. 
{\it Panel (b)}: distribution of $|\delta\mu_{TOT}|$ 
for one million of ``RGB-a-equivalent'' subsamples randomly extracted 
from the MP population. Note that the observed value lies at more than $10 
\sigma$ from the mean of the simulated samples. 
\label{conc}}    
\end{figure}

{\it What is the statistical significance of the detected difference?}
Both mono-dimensional (Fig.~3a) and bidimensional KS tests (see
Table~\ref{ks}) show that the probability that the proper motion
distribution observed for the RGB-a stars is drawn from the same
parent distribution of the MP population is lower than $10^{-13}$.  We
obtain the same answer by performing extensive Monte Carlo
simulations: $10^6$ sub-samples of the same dimension of the RGB-a
sample have been randomly extracted from the dominant (MP) proper
motion distribution. For each extracted sample the total mean proper
motion modulus $|\delta\mu_{TOT}|$ has been computed: in no case it
turned out to be larger than or equal to that observed for the RGB-a
stars.  Moreover {\em all} of the extracted sub-samples have
$|\delta\mu_{TOT}|\le0.4$, i.e., always much lower than what actually
observed for the RGB-a (Figure~\ref{conc}, Panel {\it (b)}).  These
tests put the statistical significance of the result shown in Fig.~2
beyond any possible doubt: {\em the RGB-a stars have a proper motion
distribution that is not compatible with that of the bulk (MP)
population of $\omega$~Cen}.

\citet{nor97} found evidence that, while the most metal-poor stars
in their sample ([Ca/H]$<-1.2$) rotate, the most metal-rich ones
instead show no sign of rotation. {\it Is there a connection between
this effect in the radial velocities and what we find here in the
proper motion distribution?} First of all, it must be noted that the
metal-rich sample in \citet{nor97} is mostly a subsample of our
MInt population, and it contains only a handful of stars belonging
to the RGB-a, so the two effects regard two different groups of stars.
To prove this point, we plot the metal-rich sample of Norris et al. in
Panel {\it (d)} of Figure~\ref{motion}: the difference between their
distribution and that of the RGB-a population (Panel {\it (c)} of
Figure~\ref{motion}) can be clearly appreciated. Moreover, both the
bidimensional KS tests (Table~\ref{ks}) and the boot-strap Montecarlo
simulations like those described above demonstrate that the proper
motion distribution of the RGB-a cannot be extracted from the same
parent population of the MInt, nor from the same parent distribution
of the metal-rich sample of Norris et al., at the highest level of
confidence ($P<10^{-5}$). Thus we conclude that {\em the RGB-a group
shows a proper motion distribution which is significantly different
from that of any other sub-population in $\omega$~Cen}.

On the other hand, the membership of the RGB-a stars to the
$\omega$~Cen system is proven by the radial
velocities. Table~\ref{radvel} lists the mean radial velocities for
each sub-sample defined in Section~\ref{sub-pops}.  Though radial
velocities have been measured only for ten RGB-a stars, their mean
velocity turns out to be fully consistent with that measured for the
other sub-populations. This proves that all the considered
sub-populations are locked in the same gravitational potential.

\begin{deluxetable}{lll}
\tabletypesize{\small}
\tablecaption{Bidimensional KS test for the proper motion distributions.
\label{ks}}
\tablehead{
\colhead{Sub-Populations} & \colhead{$D$} & \colhead{$P$}} 
\startdata
 RGB-a vs MP           & 0.567 & $1.6 \times 10^{-14}$ \\
 RGB-a vs N97          & 0.455 & $3.0 \times 10^{-6} $ \\ 
 MP vs HB              & 0.003 & 0.46 \\
 MInt vs \citet{nor97} & 0.13  & 0.33 \\ 
 MInt vs MP            & 0.264 & $6.5 \times 10^{-11}$ \\
 MInt vs RGB-a         & 0.396 & $5 \times 10^{-6}$ \\
\enddata
\tablecomments{Bidimensional KS test for the proper motion distributions 
of the sub-populations discussed. $D$ is the maximum difference in the
cumulative distribution, $P$ is the probability that the two
populations are drawn from the same parent distribution. }
\end{deluxetable}

\begin{deluxetable}{llll}
\tabletypesize{\small}
\tablecaption{Mean Radial velocities for the three sub-populations in 
$\omega$~Cen \label{radvel}}
\tablehead{
\colhead{Sub-Pop} & \colhead{$V_{rad}$} & \colhead{$\sigma_{V}$} & 
\colhead{N}} 
\startdata
 MP    & 234.2 & 14.5 & 186 \\
 MInt  & 231.1 & 13.0 & 76  \\
 RGB-a & 235.6 &  9.8 & 10  \\
\enddata
\tablecomments{Radial velocities are from Mayor et al (1997),
for RGB-a stars a few additional stars have been observed by
\citet{p02}.}
\end{deluxetable}

Assuming a distance of 5360~pc for the cluster \citep{ogle}, the
absolute differential proper motion of the RGB-a population
corresponds to a total mean tangential velocity of $V_t\simeq19.8$
km~s$^{-1}$   with respect to the main
system, in excellent agreement with the velocity dispersion measured
in the core of $\omega$~Centauri \citep{may97,vanl00}.  By combining
this estimate with the difference in the radial velocity between the
RGB-a and MP stars ($\Delta V_r=1.4$~km~s$^{-1}$ -- see
Table~\ref{radvel}), we obtain a total mean velocity $V_{tot}\sim20$,
still significantly lower than the half-mass escape velocity for $\omega$~Cen
($V_{esc}=44$~km~s$^{-1}$), as recently estimated by \citet{oleg}.
The observed bulk motion of the RGB-a stars is fully consistent with
being driven by the potential well of the whole cluster. We can thus
conclude that {\it the RGB-a population is a coherently moving group
of stars, gravitationally bound to the $\omega$~Cen system}.

Although a detailed discussion of the MInt properties is beyond
the scope of this paper, it is worth mentioning that also the  MInt
population shows a different proper motion distribution with respect
to the  MP population. Although the effect is smaller than the one
shown by the RGB-a population (see Figure~\ref{motion}), the tests in
Table~\ref{ks} and Figure~\ref{conc} prove that the difference is
still significant. Moreover, since this sub-population basically
coincides with the metal-rich group of \citet{nor97} (see
Table~\ref{ks}), we can confirm that {\it the global kinematical
properties of the MInt stars are distinct from those of the main
population of $\omega$~Cen.}

\section{Discussion} 

By cross-correlating the photometric catalog by \citet{p00} and the
proper motion catalog by \citet{vanl00}, we have shown that {\it (i)}
the proper motion distribution of the RGB-a population is not
compatible with the one of the dominant (MP) population; {\it (ii)}
the proper motion distribution of the RGB-a is significantly different
also from the MInt population \citep[which includes the metal-rich
group by][]{nor97}, {\it (iii)} the radial velocity ensures us that
the coherent moving RGB-a group is gravitationally tied to the main
body of $\omega$~Cen.
 
This strongly suggest that the RGB-a sub-population had an (at least
partially) independent origin with respect to the bulk of the cluster
population (i.e., MP and/or Mint stars).  This conclusion is
also supported by other observational results.  For example,
\citet{p00} showed that the spatial distribution of the RGB-a stars is
elongated perpendicularly to the major axis of the dominant MP
population of $\omega$~Centauri.
Furthermore, both high-resolution optical \citep{p02} and
medium-resolution IR \citep{o02} spectroscopic analyses have shown that
RGB-a stars are less $\alpha$-enhanced than the other sub-populations in
$\omega$~Cen. This fact suggests that the interstellar medium from which
they formed could have been polluted by SNIa ejecta, as opposed to {\em
all} the other cluster stars \cite[see, e.g.][and references
therein]{cunha}. In addition, it is worth noticing that all of the six
RGB-a stars studied before (large empty triangles in Figure 1) exhibit
strong BaII lines  \citep{evans} and in gerneral very high abundances of
post-iron peak elements \citep{evans,vws02}. 
These elements are thought to be produced mainly by
intermediate-low mass asymptotic giant brach stars \citep{busso}.

On the other hand, many authors \citep[see e.g.][and references
therein]{norris,smith} showed that the s-process elements overabundance
seems to increase continuously with [Fe/H], at least for stars belonging
to the RGB-MP and MInt populations, since only few data are available for
the RGB-a population. For example, according the abundance analysis by
\citet{vws02}, ROA300 (an RGB-a member) shows a
substantially higher abundance of s-process elements with respect to star
ROA201 (which is conversely an MInt member). If this kind of discontinuous
behaviour is confirmed, it will further support the hypothesis that the
RGB-a population had a very different evolutionary and chemical history
with respect to the other stars of $\omega$~Centauri.

In the simplest conceivable scenario, the RGB-a stars originated as an
independent stellar system, now trapped and disrupting in the
potential well of $\omega$~Centauri. Since the chance of capturing a
completely independent object orbiting in the halo of the Galaxy is
highly unlikely, it is reasonable to presume that the accreted system
was a former satellite of $\omega$~Centauri.  This fact would be in
agreement with the hypothesis that $\omega$~Centauri is in fact the
relic of a larger galaxy, partially disrupted by the Galactic tidal
field \cite[as suggested by][]{freeman,lee,dinescu,maj}.  If this is
the case, we are witnessing the process of hierarchical merging,
simultaneously occurring on two very different scales: the
$\omega$~Centauri system is merging into the Milky Way and the RGB-a
system is merging into $\omega$~Centauri.  High resolution Cold Dark
Matter (CDM) models \citep{ben} predict that hierarchical clustering
processes do occur down to the dwarf-galaxy scale ($M \sim 10^7
~M_{\odot}$) even at the present day.  Apparently, we have found an
{\em in vivo} example, suggesting that hierarchical merging occurs
also on sub-galactic scales ($M \sim 10^{4-5} ~M_{\odot}$).

Without a detailed knowledge of the structural parameters for the
RGB-a system, no quantitative conclusion can be made concerning the
timescales involved in the accretion of the RGB-a by
$\omega$~Cen. However, as an approximate criterium, we can assume that
an accreted self-gravitating stellar system can resist the tidal field
stress of the main accreting system if its density is larger than the
average density of the host system at its orbital radius
\citep{oh,andy}. From the data at our disposal \citep{p00,post,russo},
we can roughly estimate that the RGB-a population has a central
density comparable to that of the MP population and
a {\it core radius} significantly smaller    
($r_c^{RGB-a}\sim 0.5 r_c^{MP}$) than
that of the main MP population. Then, according to equation~18 by
\citet{gerh}, the RGB-a system is expected to dissolve under the tidal
strain of $\omega$~Cen if its orbital radius is lower than
$\sim1r_c^{MP}$, i.e. the ``dangerous'' zone for the accreted RGB-a
should be limited to the {\it core} of $\omega$~Cen.

The observational facts collected so far
\citep{nor97,jur,lee,p00,post} point towards a multiple merging and/or
accretion event in the past history of $\omega$~Cen. However unlikely
this may appear, we have no other way to explain the structural and
kinematical properties of the various cluster sub-populations. We may
note that, as a typical galaxy evolves by accreting smaller systems
(i.e., the Sagittarius Dwarf Spheroidal for the Milky Way), the past
history of this giant cluster (or dwarf galaxy?) was probably
characterized by multiple accreting/merging episodies. Some signatures
of these remote events are still observable in the relic of the
distrupting sub-systems.
The exact connection between the different accreted stellar
components (MInt and RGB-a), still remains to be
fully understood. A complete radial velocity study of the more metal
rich components in $\omega$~Cen will shed more light on this point.

\acknowledgments 
We thank Nicole Homeier for a critical reading of the manuscript
and an anonymous referee for helpfull suggestions which improved 
the discussion presented in the paper. This
research was partially financed by the italian Ministero
dell'Istruzione dell'Universit\`a e della Ricerca (MIUR) and by the
Agenzia Spaziale Italiana (ASI). E.P. aknowlegdes the support of the
ESO Studentship Programme.

\end{document}